# Integrating epidemiological and economic models to estimate the cost of simulated foot-and-mouth disease outbreaks in Brazil

Nicolas C. Cardenas[1], Taís C. de Menezes[2], Amanda M. Countryman[2], Francisco P.N. Lopes[3], Fernando H.S. Groff[3], Grazziane M. Rigon[3], Marcelo Gocks[3], Gustavo Machado[1,4*]

[1]Department of Population Health and Pathobiology, College of Veterinary Medicine, North Carolina State University, Raleigh, North Carolina, USA.

[2]Department of Agricultural and Resource Economics, College of Agricultural Sciences, Colorado State University, Fort Collins, Colorado, USA.

[3]Departamento de Defesa Agropecuária, Secretaria da Agricultura, Pecuária e Desenvolvimento Rural, Porto Alegre, Brazil.

[4] Center for Geospatial Analytics, North Carolina State University, Raleigh, NC, USA.

**\*Corresponding author:** gmachad@ncsu.edu.

## Abstract

The introduction of foot-and-mouth disease (FMD) leads to substantial economic impacts through animal loss, decreased livestock and meat production, increased government and private spending on control and eradication measures, and trade restrictions. This study evaluates the direct cost-effectiveness of four control and eradication scenarios of hypothetical FMD outbreaks in Rio Grande do Sul, Brazil. Our model simulation considered scenarios with depopulation of detected farms and emergency vaccination and two enhanced scenarios featuring increased capacity for emergency vaccination and depopulation. FMD outbreaks were simulated using a multi-host, single-pathogen Susceptible-Exposed-Infectious-Recovered model incorporating species-specific transmission probabilities, within-farm dynamics, and spatial transmission factors. The economic cost evaluation encompassed animal elimination (a.k.a. depopulation), carcass disposal, visits by animal health officials, laboratory testing, emergency vaccination, and sanitary barriers (a.k.a. traffic-control points), and movement restrictions due to control zones. Our results provided a range of predicted costs for a potential reintroduction of FMD ranging from $977,128 to $52,275,811. Depopulation was the most expensive, followed by local traffic control points and emergency vaccination. Our results demonstrated that higher rates of depopulation, or depopulation combined with vaccination, were the most effective strategies to reduce long-term economic impacts despite higher initial costs. Allocating more resources early in the outbreak was cost-effective in minimizing the overall effect and achieving faster eradication.

# 1. Introduction

Foot-and-mouth disease (FMD) is a highly contagious viral disease that affects animals with cloven hooves, including cattle, sheep, goats, camelids, and pigs (Moonen and Schrijver, 2000). The disease is characterized by fever, blister-like lesions, and erosion on the tongue, lips, mouth, and between the hooves. Despite the high recovery rates, FMD leads to a weakened state, weight loss, and reduced milk and meat production (Azeem, 2020). Outbreaks in FMD-free countries cause yearly losses of over USD 1.5 billion (Knight-Jones and Rushton, 2013). Trade restrictions due to FMD occurrence limit access to profitable international markets for livestock and animal products (James and Rushton, 2002). Additionally, in FMD-free countries, outbreaks lead to significant costs for countries to regain disease-free status (James and Rushton, 2002). For instance, a large multi-state FMD outbreak in Australia could lead to costs ranging from $60

million to AUD 373 million, with significant additional costs for compensating farmers (Benjamin et al., 2024). Past large epidemics like the 2001 event in the United Kingdom and The Netherlands led to the culling of more than 6 million animals. The loss to the agricultural sector was approximately €3.2 billion, with additional costs to other industries, such as tourism, with the sum of expenditure estimated between €2.7 billion and €3.2 billion (Bouma et al., 2003; Thompson et al., 2002). Similarly, the average cost estimates for short-term effects of FMD in Austria are approximately €543 million, reaching as high as €1.29 billion in a worst-case scenario (Marschik et al., 2021b, 2021a).

      Rich and Winter-Nelson (2007) analyzed the economic impact of FMD in Argentina, Uruguay, and Paraguay and demonstrated losses associated with FMD outbreaks using market values with the dynamic and spatial epidemiological model (Rich and Winter-Nelson, 2007). Countryman and Hagerman (2017) highlighted that the 2001 FMD outbreaks in Brazil, Argentina, and Uruguay caused significant short-term disruptions in livestock production. These disruptions were primarily driven by the countries' heavy reliance on beef exports, as well as the effects of depopulation and interruptions in beef production (Countryman and Hagerman, 2017). Various response strategies were implemented, including regionalization, vaccination, and culling (PANAFTOSA-OPS/OMS, 2021). Brazil and Argentina used regionalization to maintain exports from FMD-free areas, while Uruguay combined culling with emergency vaccination (Rich and Winter-Nelson, 2007). These strategies incurred significant economic costs, including direct eradication expenses and trade losses (Countryman and Hagerman, 2017). Particularly for Uruguay, the 2001 FMD outbreak had a considerable financial impact, with estimated losses of $730 million and a 1.9% reduction in Gross Domestic Product between 2001 and 2003 (PAHO, 2020). De Menezes et al. (2023) demonstrate that the long-term economic impacts of potential

FMD outbreaks in Brazil extend beyond eradication and direct costs. The countermeasures required to control and eradicate an FMD outbreak result in expected prolonged recovery times as herds are rebuilt and market access is regained  (Abao et al., 2014; De Menezes et al., 2023).

The most recent FMD outbreaks in the Americas occurred in Colombia in 2017 and 2018. In 2017, 36 farms were infected, while in 2018, five infected locations were identified, leading to the quarantine of 132,752 animals across seven municipalities (PANAFTOSA et al., 2021). Understanding the potential introduction of FMD and the associated costs of outbreak control is crucial for supporting contingency plans and preparing effective strategies.

Mathematical modeling studies are frequently employed to inform or evaluate policy effects (Brauer, 2017; Marschik et al., 2021b; Rivera et al., 2023). In recent years, such models have become integral to assessing the effectiveness of key FMD countermeasures, including reactive vaccination, culling, and movement restrictions (Cabezas et al., 2021; Marschik et al., 2021b; Perez et al., 2004). Here, mechanistic simulation models enable policymakers to explore hypothetical scenarios, estimate economic losses under varying outbreak conditions, and evaluate control strategies. By modeling disease spread patterns, mathematical models predict epidemic trajectories, measure the impact of control measures, and estimate associated economic costs (Kretzschmar and Wallinga, 2009), offering critical insights into how FMD may spread through livestock populations.

This study utilizes results from FMD outbreak simulations based on a multi-host, single-pathogen Susceptible-Exposed-Infectious-Recovered (SEIR) model (Cespedes Cardenas et al., 2024) to benchmark the costs of four control and eradication scenarios for simulated outbreaks in Rio Grande do Sul, Brazil. The analysis focuses on estimating the direct costs of short-term outbreaks and evaluating the cost-effectiveness of different control measures.

## 2. Material and methods

### 2.1. Datasets

*Population data*

We used the Agricultural Defense System (SDA) (SDA, 2024) dataset comprising animal movement records from 355,676 farms in Rio Grande do Sul, Brazil, including cattle, buffalo, swine, sheep, and goats. After excluding 70,853 farms (19.94%) without geographical coordinates, information on the number of animals, or no movement records between August 24, 2022, and August 24, 2023, our final dataset comprised 284,823 farms. Cattle and buffalo farms were grouped into "bovines," and sheep and goats aggregated into "small ruminants."

*Birth and death data*

In Brazil, producers are required by law to update their population data once a year. Failure to report prevents producers from requesting movement permits required to move animals, and they are also penalized financially. Here, we use the population data to represent the total number of animals on each farm, which consisted of 273,787 births and 268,790 deaths reported by producers, detailing the total number of animals born alive and deaths due to natural or disease causes. More details about the population data and distribution are shown elsewhere (Cespedes Cardenas et al., 2024).

*Animal movement data*

The SDA dataset also stored 763,448 unique animal movements from August 24, 2022, to August 24, 2023, representing both between-farm and farm-to-slaughterhouse movements. After removing 106,481 records (13.9%) due to missing information, including movements without the number of animals transported, identical origin and destination, or movements involving unregistered or out-of-state premises, we end up with 413,939 between-farm movements and 243,028 farm-to-slaughterhouse movements were included in the database for this study.

## 2.2 SEIR model description

A multi-host, single-pathogen, coupled multiscale model, implemented in our *MHASpread* model (Cespedes Cardenas and Machado, 2024) and described in (Cespedes Cardenas et al., 2024), was employed to simulate FMD outbreaks. Our model integrates within-farm dynamics, including birth and death rates and between-farm transmission, accounting for species-specific transmission probabilities, latency, and infectious periods. Spatial transmission uses a kernel transmission approach, with the transmission probability decreasing with distance. Simulations began with an initial seeding of FMD into farms, followed by a silent spread period of ten days before control actions were applied. A baseline scenario from now so-called *base scenario* considered depopulation, emergency vaccination, movement standstill (in all control zones: infected, buffer, and surveillance), and control zones, in which a daily vaccination of 15 farms and depopulation of four farms. Three alternative scenarios consisted of variations from the *base* scenario via increasing capacities for vaccination and/or depopulation. Thus, the alternative scenarios were double vaccination and depopulation named *base x2*, triple vaccination and depopulation named *base x3*, and a scenario consisting of triple depopulation capacity of the *base* scenario named *depopulation*. In Supplementary Material Table S1, we compare each scenario's a) initial conditions, detection dynamics, movement restrictions, depopulation conditions, and how vaccination was applied. The model results ultimately allowed for the analysis of these strategies' effectiveness and itemized costs.

## 2.3 Economic model description

We used Brazilian currency (Real-R$) to collect costs, later converted to United States dollars using an exchange rate of 1 BRL to 0.18 USD ($) as of 18th of July 2024 (Banco Central do Brasil, 2024).

*Depopulation cost*

Each farm's total number of animals determined the depopulation cost, calculated in each simulation and by farm, day, and scenario. The average animal cost is based on the 2023 average price for each species. The cattle average price considered the prices of cattle sold for slaughter in Rio Grande do Sul (CEPEA, 2024a), the average swine price was calculated from the Cepea/ESALQ live swine Index for Rio Grande do Sul, and the lamb price in Rio Grande do Sul was used as a proxy for the average price of small ruminants (CEPEA, 2024a).

According to Brazilian legislation, indemnification of culled animals due to FMD has a percentage cap of 50% of the total value of the animal (MAPA, 2020). For bovines, the cost per animal loss after the 50% deduction for indemnification is $412.32; for swine, it is $51.85; and for small ruminants, the animal loss corresponds to $26.09 per head. The distribution of the number of culled animals by species for each scenario is shown in Supplementary Material Figure S1, and the average cost of animal loss in each scenario is presented in Supplementary Material Figure S2. In the most recent FMD outbreaks in the Rio Grande do Sul, culling was done only via "sanitary rifle," and therefore, we include the cost for ammunition, which equals $1.26 per head, regardless of the species. The average depopulation cost in each scenario is presented in Supplementary Material Figure S3. The total depopulation cost ($Depop$) is calculated following the equation below:

$$Depop_{i,t} = animalsculled_{i,t} * animalcost_{i,t,sp} * munitioncost_{i,t}$$

Where the cost in each simulation $i$ per day $t$ is calculated considering the number of animals culled ($animalsculled$), the cost of compensation for the total animals (culled $animalscost$ in a day $t$ according to the species $sp$ (bovine, swine, or small ruminant), and the ammunition cost per head ($munitioncost$).

*Carcass disposal*

To determine the cost of carcass disposal, we considered the average daily cost of the use per rented excavator machine per affected farm and disposal capacity in terms of the number of animals. The cost of carcass disposal ($carcass$) was calculated as follows:

$$Carcass_{i,t} = capacity_{sc} * excavator_{i,t}$$

The capacity of carcass disposal varies according to each simulated scenario ($sc$). In the *base* scenario, four farms were depopulated daily. Eight farms were depopulated daily in the *base x2* scenario, and 12 were depopulated daily in the *base x3* and *depopulation* scenarios. Supplementary Material Figure S4 presents the average carcass disposal cost in each scenario.

*Official staff visits*

The cost per visit by animal health officials was calculated based on the average daily salary of $13.53. We assumed that for each farm visit, two animal health officials were needed and that each infected farm requires ten visits to identify the infected farm and eliminate all animals. The salary cost is calculated as follows:

$$Officialstaffvisits_{i,t} = detectedfarms_{i,t} * paymentinspector * ninspector * nvisifarm$$

Where $detectedfarms_{i,t}$ is the number of detected farms for simulation $i$ in the time $t$, $ninspector = 2$, and $nvisitfarm = 10$. Once a farm is detected, no additional costs are considered for the following days, assuming that the two animal health officers will complete their ten visits. Supplementary Material Figure S5 presents the average carcass disposal cost in each scenario.

*Laboratory testing*

Each detected FMD infection triggers control areas, which, for Brazil (MAPA, 2020), requires forming an infected zone with a three-km radius, followed by a buffer zone of seven and 15 km surveillance zone. For this study, we assumed that once these areas are created, they remain active until the end of the simulation (Supplementary Material Figure S6). Samples are collected from farms within a ten-km radius of detected infected farm(s) within the limits of infected and buffer zones. The model calculated the number of farms tested in both zones based on a prevalence of 50%, margin of error of 5%, and a population size based on the total number of farms in the infected and buffer zones. The population size is determined by the number of farms in infected and buffer zones on the last day of each simulation; therefore, it is calculated and added to the cost of the simulation on the last day of the outbreaks. The distribution of the number of sampled farms per scenario is presented in Supplementary Material Figure S7. The cost of laboratory tests was calculated based on the total number of sampled farms in the sampled zones for each simulation, corresponding to the accumulated costs on the final day of the simulation. The total cost of laboratory tests was calculated using the following formula:

$$Labtestcost = sampledfarms_{i,ft} * (csample + cshipping + PPE)$$

Where $sampledfarms_{i,ft}$ represents the number of sampled farms in each simulation $i$ on the last day of simulations $ft$. $csample$ represents the cost of the laboratory test, which is estimated, on average, to be $9.90. The estimated shipping cost per sample $cshipping$ is equal to $0.20. $PPE$ includes the necessary supplies to collect the sample (e.g., disposable clothing, needles, syringes) for $6.30 per sample, according to the diagnostic laboratory accredited by the

Brazilian government (MAPA, 2024). The cost of a laboratory test for each scenario is presented in Supplementary Material Figure S8.

*Reactive vaccination*

A seven-day delay was applied at the beginning of the vaccination process to account for the time required to prepare and distribute the vaccine. The estimated cost of an FMD vaccine dose is $0.65 at 2023 prices, based on the study of (Miranda S.H.G. et al., 2018), and we assume that each animal would be vaccinated twice. The vaccination cost was calculated using the following formula:

$$Costvacc_{i,t} = bovvacc_{i,t} * pvaccine$$

Where $bovvacc$ represents the number of bovines vaccinated for simulation $i$ at time $t$ and $pvaccine$ corresponds to the price of the vaccine dose. The vaccination cost for each scenario is presented in Supplementary Material Figure S9.

*Traffic control points*

The cost of traffic control points is based on the perimeter size of the control zones and the number of roads within affected areas. The 25 km perimeter set around detected infected farms, including all under surveillance (Brasil. Ministério da Agricultura, 2020). Under the national control plan, all movements in or out of the 25 km are prohibited and monitored for 30 days following the last detected case (MAPA, 2020). This requires establishing control access points to regulate entry and exit from control zones. To determine the number of control access points needed, we used regional road maps (SAPDR-RS, 2024) and overlaid them with the control zone to determine the required number of control access points (Figure 1). This interpolation determined the number of access points needed to regulate animal movements. Figure 1 illustrates how the number of control access points is determined, and the distribution of the control access points is provided in Supplementary Material Figure S10.

Figure 1

We established the average cost of each traffic control point, including personnel and mileage costs. The total cost of local control points is calculated as follows:

$$Controlpoints_{it} = barriercostday * numaccespoints_{i,t}$$

Where $barriercostday$ is the cost of traffic control point, and $numaaccesppoints_{i,t}$ what is the number of traffic control points for the simulation $i$ and the day $t$.

Altogether, items that compose the cost calculations and corresponding sources are described in Table 1. The total cost of each simulation of control actions was calculated as follows:

$$Totalcost = Depop_{i,t} + Carcass_{i,t} + Officialstaffvisits_{i,t} + Labtestcost + Costvacc_{i,t} + Controlpoints_{it}$$

## 2.4 Software

This study was conducted in R statistical software v. 4.1.1 (R Core Team, 2024) using the following packages: tidyverse (Wickham et al., 2019) and sf (Pebesma, 2018).

## 3. Results

*Initial spread and performance of the control actions*

We simulated 7,010 outbreaks across four scenarios: *base*, *base x2*, *base x3*, and *depopulation*. Each scenario started with identical initial conditions and the start of the implemented control (Supplementary Material Table S1), resulting in a median of 25 infected farms on day 25 (Figure 2). The *base* scenario exhibited a gradual decline in prevalence, reaching zero by day 109, with a median of 18 infected farms. The *base x2* scenario showed a steeper epidemic decline, achieving zero infected farms by 51 days and a median of 16 infected farms. The *base x3* and *depopulation* scenarios eliminated all outbreaks by day 43 and day 45, respectively. The *base x3* scenario had a median of 14 infected farms, while the *depopulation* scenario had a mean of 18.7 and a median of 14 infected farms.

Figure 2

*Total FMD outbreak cost*

Table 2 presents the median final cost and the duration of control actions for each scenario on the last day of the outbreak. This refers to the simulation results from the last day of control action in each simulation, regardless of whether the outbreak was contained. For example, in some simulations, the final day might be day five, while in others, it could be day 56. In order of effectiveness and cost, in the *base x3* scenario, the median cost is $1,535,671, and the actions last a median of 15 days. The *depopulation* scenario has a similar median cost of $1,552,303, with the duration of actions lasting a median of 15 days. The *base x2* scenario sees a median cost increase to $1,661,625, and the control actions extend to a median of 17 days. Finally, the *base* scenario incurs the highest median cost of $1,888,969, with control actions lasting a median of 23 days.

Figure 3 displays the predicted cost; for descriptive purposes, we divided the timeframe into three outbreak phases: from day 20 to 25, from day 25 to 50, and from day 50 to 100, describing the median values by scenarios. During the initial outbreak phase from days 20 to 25, all scenarios show a steep increase in cost due to the initial impact of interventions being applied.

The *depopulation* scenario peaks first, reaching a median of $3,173,378, followed closely by *base x2* at approximately $3,138,006 and *base x3* at about $2,943,841. In the middle outbreak phase, going from day 25 until day 50, the base and *base x2* scenarios continue to increase linearly, reaching around $3,279,695 and $6,197,508, respectively, without fluctuations, while *depopulation* and *base x3* show a reduction in their rate of increase after their initial peaks, stabilizing or decreasing to approximately $3,313,208 and $3,016,134, respectively. In the final outbreak phase from day 50 to day 100, the *base* scenario maintains a steady linear increase, reaching approximately $9,482,079. Meanwhile, the *base x2* scenario stabilizes after the initial decline, with costs rising again to approximately $13,319,857.

Figure 3

The cost analysis reveals a notable expense reduction across all scenarios compared to the *base* scenario. The *base x3* scenario achieved the highest cost reduction, decreasing expenses by a median of 18.71% (IQR: 20.44% to 10.79%). The *depopulation* scenario also showed significant savings, with a 17.82% (IQR: 19.37% to 9.77%) reduction relative to the *base* scenario. The *base x2* scenario, while still effective, resulted in a 12.04% (IQR: 14.33% to 6.35%) cost reduction compared to the *base* scenario.

Figure 4 shows the costs associated with each economic parameter used to calculate the total cost across four scenarios on the final day of each simulation. The *base* scenario consistently exhibits significantly higher costs and varied eradication periods across most items. Notably, depopulation reached a median of $1.77 million. Interestingly, traffic-control points

were the largest compared to other scenarios, reaching $146,422, followed by carcass disposal costs with a median of $21,633.

Figure 4

For the *base x2* scenario, the median depopulation and carcass disposal costs were $1.6 million and $21,633, respectively. The emergency vaccination also incurs considerable expenses, with costs of $110,167, followed by local control points and carcass disposal, at $104,545 and $21,633, respectively.

In the *base x3* scenario, the median depopulation cost was $1.51 million, with traffic-control points costs increasing to $94,744. Carcass disposal costs remained $27,041, while emergency vaccination costs were $101,423. For the *depopulation* scenario, laboratory testing had a median cost of $6,232, while carcass disposal reached $27,041. Official staff visits incurred lower expenses, with a median of $1,624, while local control points presented substantial costs, with a median of $95,635. Depopulation was the most expensive component, with a median cost of $1.54 million. The cost of all variables over all parameters is presented in Supplementary Material Table S2. The proportion of the cost considering the total is presented in Supplementary Material Table S2 and Supplementary Material Figure S11.

## 4. Discussion

This study examined the epidemiological and economic implications of FMD reintroduced in Rio Grande do Sul, Brazil, using a multi-host SEIR model (Cespedes Cardenas et al., 2024) and economic cost across four control strategy scenarios. Most FMD economic studies focus on cost in endemic areas (Baluka, 2016; Chaters et al., 2018; Compston et al., 2022; Nampanya et al.,

2016; Shankar et al., 2012; Sinkala et al., 2014; Tadesse et al., 2020; Truong et al., 2018; Young et al., 2016) or the potential long-term economic impacts of FMD in disease-free regions (Australian Bureau of Agricultural and Resource Economics and Sciences (ABARES), 2019; Cairns et al., 2017; Feng et al., 2017; Junker et al., 2009; Marschik et al., 2021b; Tozer and Marsh, 2012; Wittwer, 2024). A few studies examine the direct short-term costs in FMD-free regions, especially in terms of cost comparison between different control strategies (Australian Bureau of Agricultural and Resource Economics and Sciences (ABARES), 2019; Benjamin et al., 2024; Elbakidze et al., 2008; Garner and Beckett, 2005; Halasa et al., 2020; Hayama et al., 2017; Lyons et al., 2021; Marschik et al., 2021b; Pesciaroli et al., 2025; Waret-Szkuta et al., 2017; Wittwer, 2024). Our study estimates the total costs but also provides expenses of itemized direct costs, including depopulation, carcass disposal, laboratory testing, emergency vaccination, and surveillance, which encompasses the costs of farm visits by animal health officials and monitoring at traffic control points. Notably, this study is one of the few that conducts such estimations in Latin America (De Menezes et al., 2023, 2022).

With triple the Rio Grande do Sul's capacities for vaccination and depopulation (*base x3*) and a scenario with triple the state's capacity (*depopulation*) but no emergency vaccination eliminated all outbreaks by days 43 and 45, respectively. In contrast, the *base* scenario stamped out all outbreaks by day 109, while double the state's capacity for depopulation and vaccination (*base x2*) achieved a steeper decline, reaching zero infected farms by day 51. The primary factors in reducing the time and number of infected farms were changes in depopulation and vaccination rates. Similar findings were observed in Japan, Italy, and Austria, where depopulation was identified as the most effective and efficient control strategy, particularly in the northern regions (Hayama et al., 2017; Marschik et al., 2021b; Pesciaroli et al., 2025).

Over 100 days of control actions, cost patterns varied across scenarios. In the early phase of the simulated epidemics, all scenarios exhibited a sharp cost increase, with the *depopulation* scenario peaking at $3.71 million, the *base x2* scenario at $3.13 million, and the *base x3* scenario at $2.94 million. In the middle outbreak phase, costs for the *base* scenario continued to rise, reaching $3.27 million. In comparison, costs for the *base x2* and *base x3* scenarios either stabilized or decreased at $6.19 million, and the *base x3* and *depopulation* scenarios decreased to $3.01 million and $3.31 million, respectively. By the final outbreak phase, the costs for the *depopulation*, *base*, and *base x2* scenarios rose significantly. We found that while depopulation strategies led to higher initial costs at the onset of outbreaks, it was the most effective in reducing epidemic incidence. The higher costs early on are due to the more significant number of infected farms, leading to more animals needing to be culled, more control zones being established, and large sums of animals being vaccinated. However, as the is decreases, the number of animals requiring vaccination and culling decreases, lowering the cost. This trend is reflected in the median costs, which drop from $1,888,969 in the *base* scenario to $1,535,671 in the *base x3* scenario, a reduction of $353,298 (18.7%). Similarly, in the *depopulation* scenario, median costs decrease to $1,552,303, a reduction of $336,666 (17.8%). Additionally, the median duration of control measures shortens from 23 days in the *base* scenario to 15 days in both the *base x3* and *depopulation* scenarios, representing an 8-day reduction (34.8%). These results are consistent with other studies, where costs for prompt culling and early detection scenarios represented 30% and 2% of the *base* scenario's costs, respectively (Hayama et al., 2013; Hayama et al., 2017). The findings underscore the critical importance of detecting infected farms as early as possible and culling animals promptly (Marschik et al., 2021a, 2021b). Overall, managing simulated outbreaks in Rio Grande do Sul was less expensive compared to recent simulation studies in other countries, including Austria, where the cost ranged from €219 to €1,289 million, with 4% (1–7%) being direct costs (Marschik et al., 2021c).

For our study area, we demonstrated that depopulation accounted for more than 86% of the total costs across all scenarios. Thus, exploring alternative strategies that reduce the need for extensive culling, such as targeted vaccination or improved biosecurity measures, could be beneficial in improving FMD control costs. Our study aligns with findings from Denmark, where depopulation in zones around infected herds (infected zone) was identified as the most cost-effective strategy for controlling the disease (Boklund et al., 2009), similar to Sweden, where vaccination did not lower the expected number of infected herds, while preemptive depopulation reduced infections in extreme scenarios but doubled the number of herds culled (Dórea et al., 2017).

In contrast to our findings, surveillance associated with control zones in the Austria modeling work was the most expensive item, with an estimated outbreak cost between €269 and €581 million, and the depopulation scenario specifically costing around €460 million (€221–€853 million) (Marschik et al., 2021c, 2021a). However, the vaccination strategy in Austria yielded similar results to the reference scenario regarding infected farms and total outbreak control costs (Marschik et al., 2021c, 2021a). In Pakistan, the most significant cost was vaccination, accounting for 56% of total expenses (Lyons et al., 2021). These results demonstrate the heterogeneity of outcomes across countries and suggest that costs vary according to population dynamics (Ward and Perez, 2004), population densities (Meadows et al., 2018; Seibel et al., 2024), animal movement patterns (Contina et al., 2024), and differing assumptions across modeling frameworks. Additionally, compensation costs for depopulation activities can rise substantially when other species are affected, especially in regions with significant swine production (Machado et al., 2021). Our findings demonstrate that although enhanced control strategies may require higher initial investments, they are more effective in reducing the long-term direct cost of stamping out FMD outbreaks. Rapid and aggressive measures, particularly those involving higher rates of depopulation and depopulation combined with vaccination, are crucial for minimizing costs and achieving faster disease eradication. Thus, allocating more resources at the onset of an outbreak is a more cost-effective approach to controlling and eradicating FMD. The strain on public funds highlights the need to continue supporting emergency fund reserves such as the Fundo de Desenvolvimento e Defesa Sanatiria Animal (FUNDESA), which aims to complement actions for animal health development and defense in the state of Rio Grande do Sul. Fostering such funds could be pivotal in ensuring that resources are readily available to respond quickly and effectively to FMD outbreaks, thereby mitigating the economic and health impacts on the agricultural sector.

## 5. Limitations and further remarks

This study did not consider topographical and environmental features contributing to FMD airborne spread (Brown et al., 2022). The costs associated with using excavators, animal health official salaries, sampling (including shipping), personal protective equipment, and traffic control points were approximated due to the lack of data from past outbreaks, introducing potential uncertainties in the results. Other direct and indirect costs, such as personal expenses incurred by producers and the economic impact of export restrictions, were also not considered.

The sampling decision for surveillance followed the Brazilian FMD response plan (MAPA, 2020), which provided guidance for sampling and diagnostic tests but did not specify the areas or number of animals to be sampled. Therefore, in collaboration with the state's animal health officials, we implemented random farm sampling for the required control zones within the state's capacity. The selection of farms in control zones was determined on a case-by-case basis (animal health official personal communication, 2024); for instance, only farms identified with clinical signs of FMD are sampled, only farms with clinical signs of FMD, those directly linked to infected farms, or those near suspect farms were sampled. Our results should be interpreted as an approximation rather than a precise cost estimation for the future FMD reintroduction in Rio Grande do Sul, Brazil. However, our results provide a range and maximum estimated costs for several control and elimination scenarios. Ultimately, they help calculate the overall and itemized cost difference among the scenarios. Decision-makers shall recognize both the limitations and benefits of these models when using them to inform policy decisions. Future work will address long-term economic losses, including the impacts on local and international markets.

## 6. Conclusion

This study estimated the direct cost of eliminating an FMD outbreak in a dense livestock state in Brazil. Our modeling simulated the current state-level response plan, including movement standstill, control zone restrictions and testing, depopulation, and emergency vaccination, resulting in a median cost of $1.89 million and 23 days until all outbreaks were eliminated. The best-performing simulated control scenario was tripling the state's depopulation capacity, which would have reduced the median cost to $1.54 million and shortened the median outbreak duration to 15 days, achieving an 18.7% cost reduction and eliminating the outbreak 8 days sooner. Our modeling work ranked the total cost of a range of control actions, demonstrating that depopulation had a significant financial impact, with costs varying across scenarios.

## Ethical statement

The authors confirm the journal's ethical policies, as noted on the journal's author guidelines page. Since this work did not involve animal sampling or questionnaire data collection by the researchers, ethics permits were not needed.

## Data Availability Statement

The data that support the findings of this study are not publicly available and are protected by confidential agreements. Therefore, they are not available.

**List of tables**

**Table 1.** Parameters and costs are used to estimate the economic impact of the outbreak.

| Item | Cost | Description | Source |
|---|---|---|---|
| Cattle | $412.32 per animal | Average cattle price in 2023 for estimation of animal losses | (CEPEA, 2024b) |
| Swine | $51.85 per animal | Average swine price in 2023for estimation of animal losses | (CEPEA, 2024b) |
| Small ruminants | $26.09 per animal | Average lamb price in 2023 (a proxy for the price of small ruminants) for estimation of animal losses | (CEPEA, 2024b) |
| Munition | $1.26 per animal | Cost of ammunition per animal culled | (Receita Federal do Brasil, 2023) |
| Excavator | $450.68 per day | Cost of the excavator used to bury animals | Personal communications health officials of the State of Rio Grande do Sul (Fernando Henrique Sauter Groff, Grazziane Maciel Rigon, and Marcelo Gocks) |
| Animal health official salary | $13.53 daily salary per person | Salary of the official veterinary service staff who work on disease control and eradication actions | |
| Laboratory test | $9.90 per animal | Cost to perform a diagnostic test for FMD | Secretariat of Agriculture and Supply of the State of São Paulo/Biological Institute – Animal and Plant Health Center (Sao Paulo, 2024) |
| Sample (shipping) | $0.20 per animal | Cost of shipping one sample to the diagnostic laboratory | Personal communications health officials of the State of Rio Grande do Sul (Fernando Henrique Sauter Groff, Grazziane Maciel Rigon, and Marcelo Gocks) |
| Personal protective equipment (PPE) | $6.30 per animal | Necessary materials to collect the sample (e.g., disposable clothes, needles, syringes | |
| Vaccine dose | $0.65 per animal | Cost of vaccine shot per head | Miranda et al. (2018) |
| Traffic-control point | $54.69 per point/daily | Cost to establish a traffic-control point control that includes the barrier and mileage | Personal communications health officials of the State of Rio Grande do Sul |

|  |  |  | (Fernando Henrique Sauter Groff, Grazziane Maciel Rigon, and Marcelo Gocks) |
|--|--|--|--|

**Table 2.** Distribution of the cost at the end of each simulation by scenario.

| Scenario | Median, interquartile range (IQR), and maximum cost | Days working in control actions |
|---|---|---|
| Base | $1,888,969 (IQR: $977,128 - $3,241,477, max: $21,478,453) | 23 (IQR: 18 - 30), max: 109 |
| Base x2 | $1,661,625 (IQR: $837,047 - $3,035,568, max: $20,655,679) | 17 (IQR: 14 - 20), max: 52 |
| Base x3 | $1,535,671 (IQR: $777,361 - $2,891,869, max: $20,587,796 | 15 (IQR: 14 - 17), max: 40 |
| Depopulation | $1,552,303 (IQR: $787,844 - $2,924,621, max: $52,275,811) | 15 (IQR: 14 - 17), max: 45 |

**List of Figures**

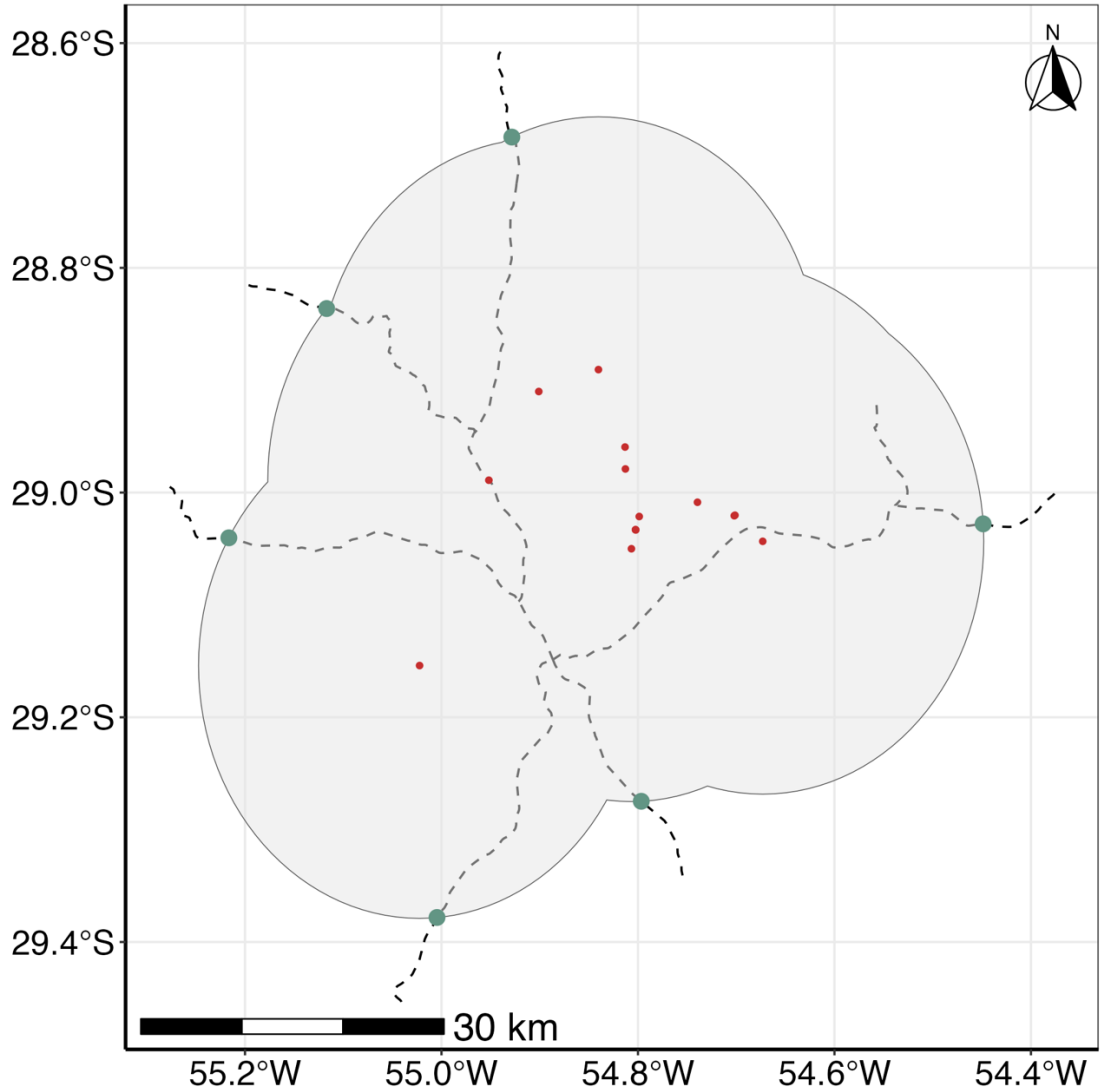

**Figure 1. Traffic control points.** The green dots represent traffic control access points established because of the intersection between the border of the control area and rounds. The dashed lines indicate the roads, while the red dots denote the detected infected farms.

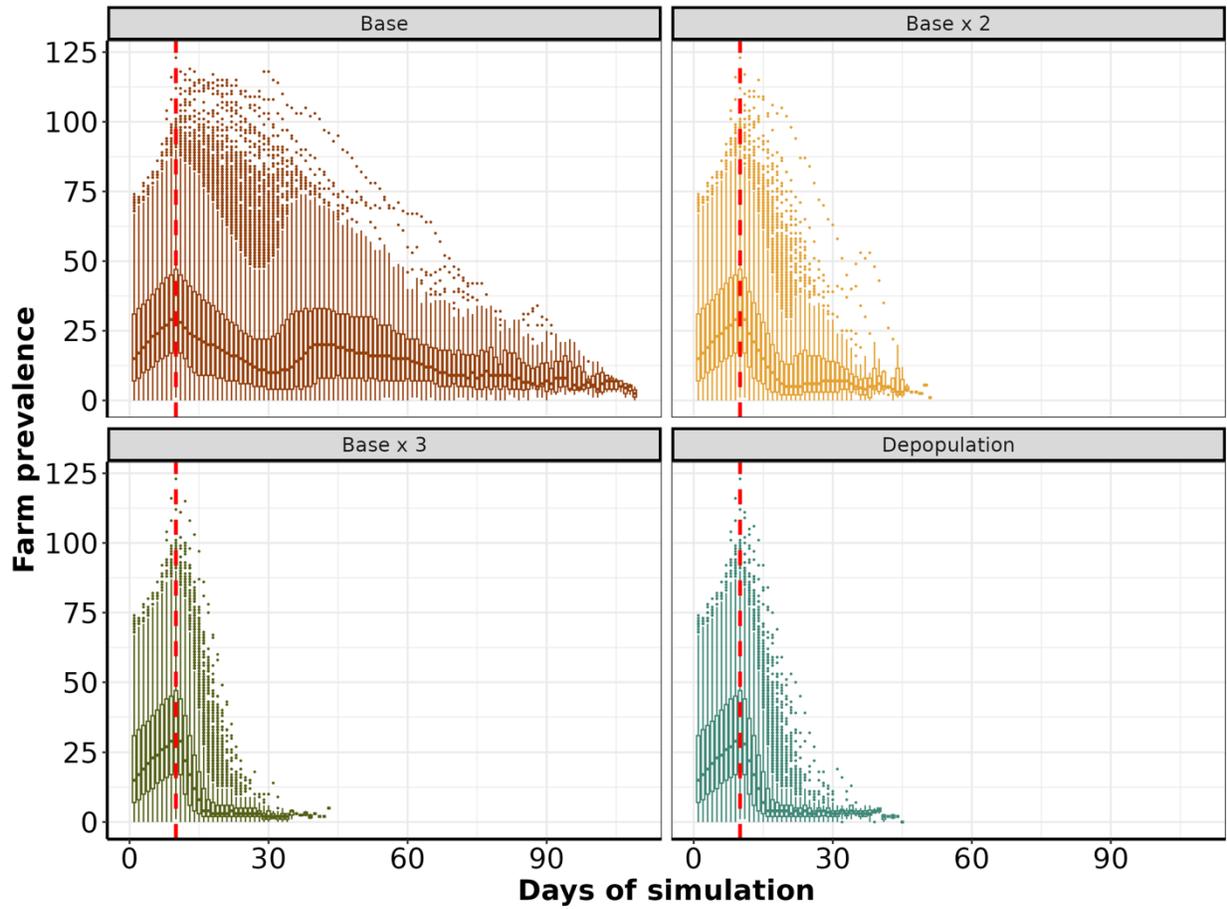

**Figure 2. Box plot of epidemic trajectories**. The y-axis denotes the number of infected farms, while the x-axis represents the simulation days. The red dashed line represents the beginning of control actions.

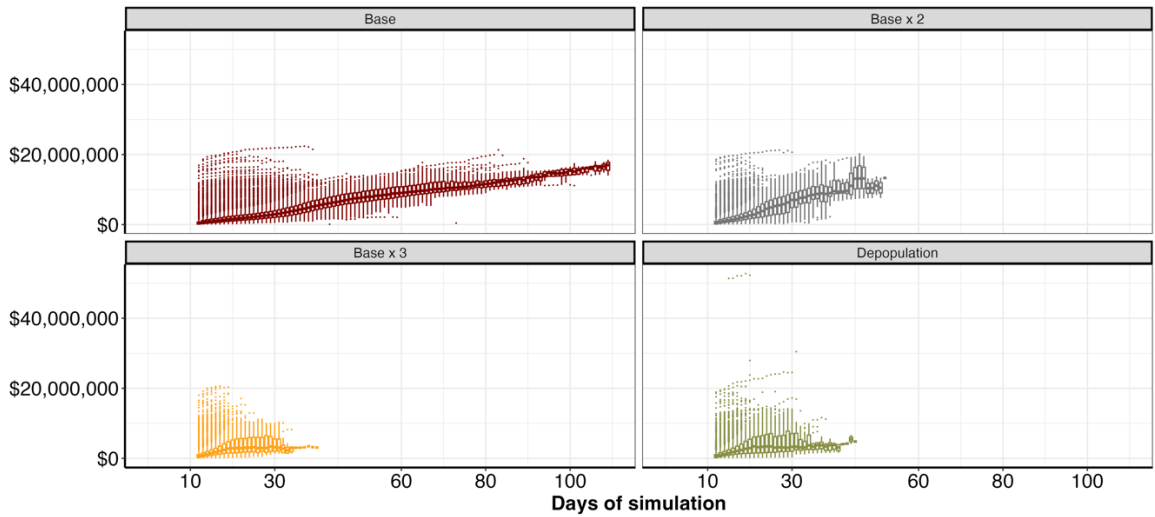

**Figure 3. Daily cost for each scenario.** The total outbreak cost is from the initial control date on day 11 until the last day of control actions or complete elimination of outbreaks.

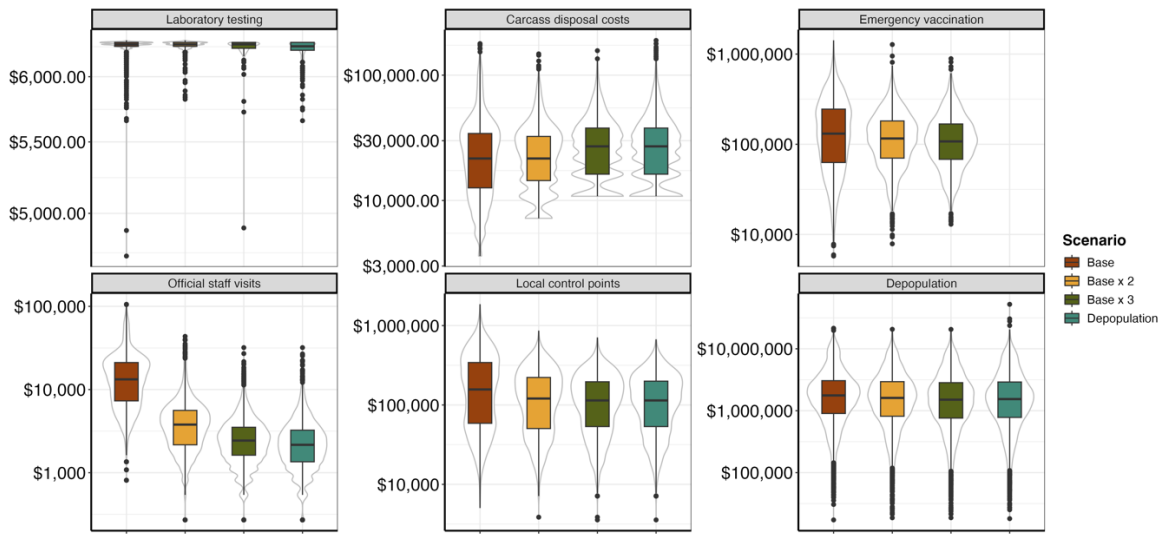

**Figure 4. Itemized costs by scenario**. The box plot represents the quantiles, while the violin in the background represents the number of observations in each value. The colors present the scenarios, and the y-axis is the cost; the distance between values is in the log10 scale.

# Supplementary Material

**Integrating epidemiological and economic models to estimate the cost of simulated foot-and-mouth disease outbreaks in Brazil**

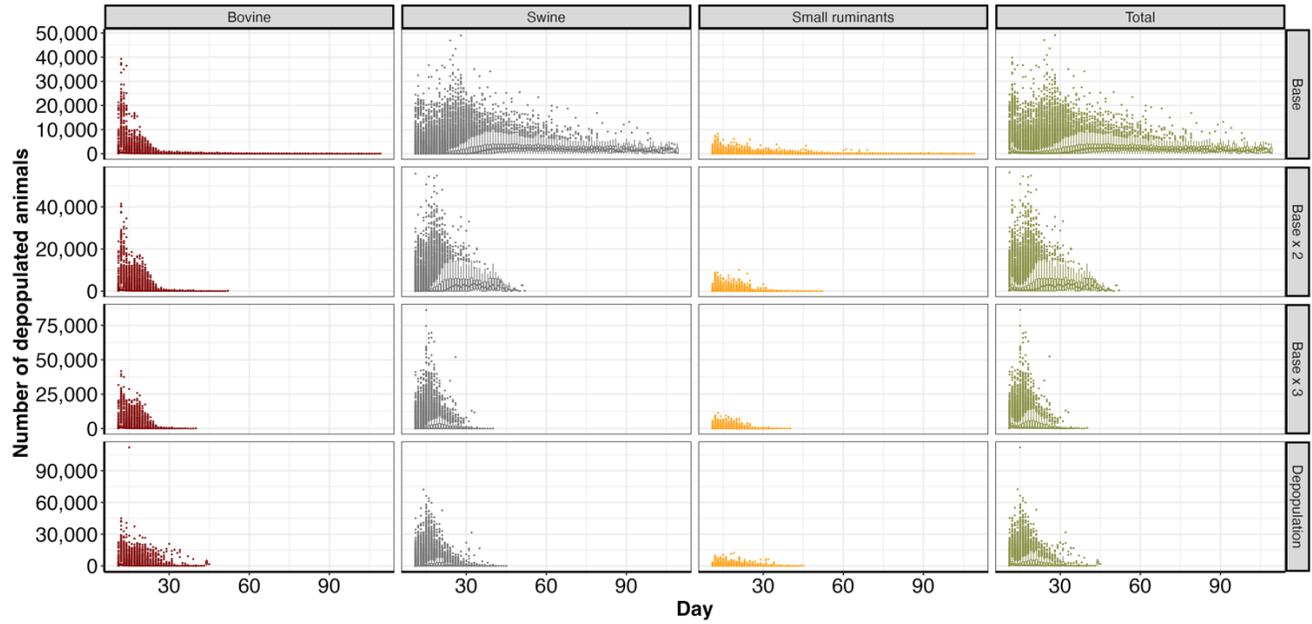

**Supplementary Material Figure S1.** The distribution of culled animals by species for each scenario.

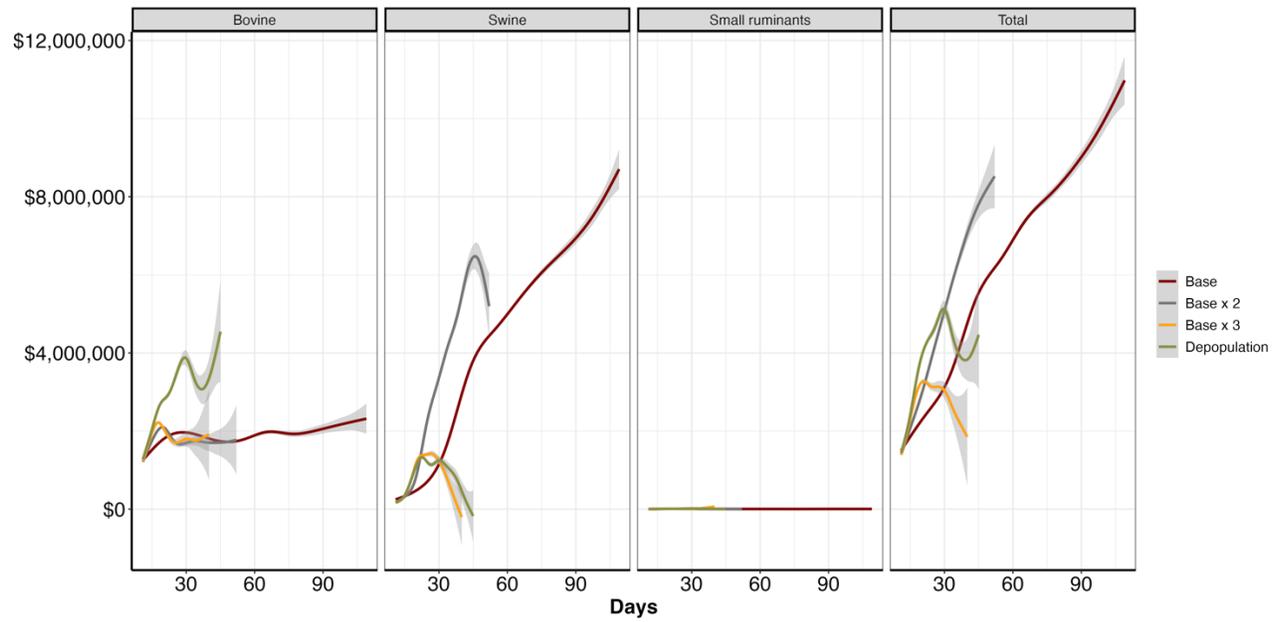

**Supplementary Material Figure S2.** The cumulative cost of the culled animals by species for each scenario. The lines are fitted based on a Generalized additive model (GAM).

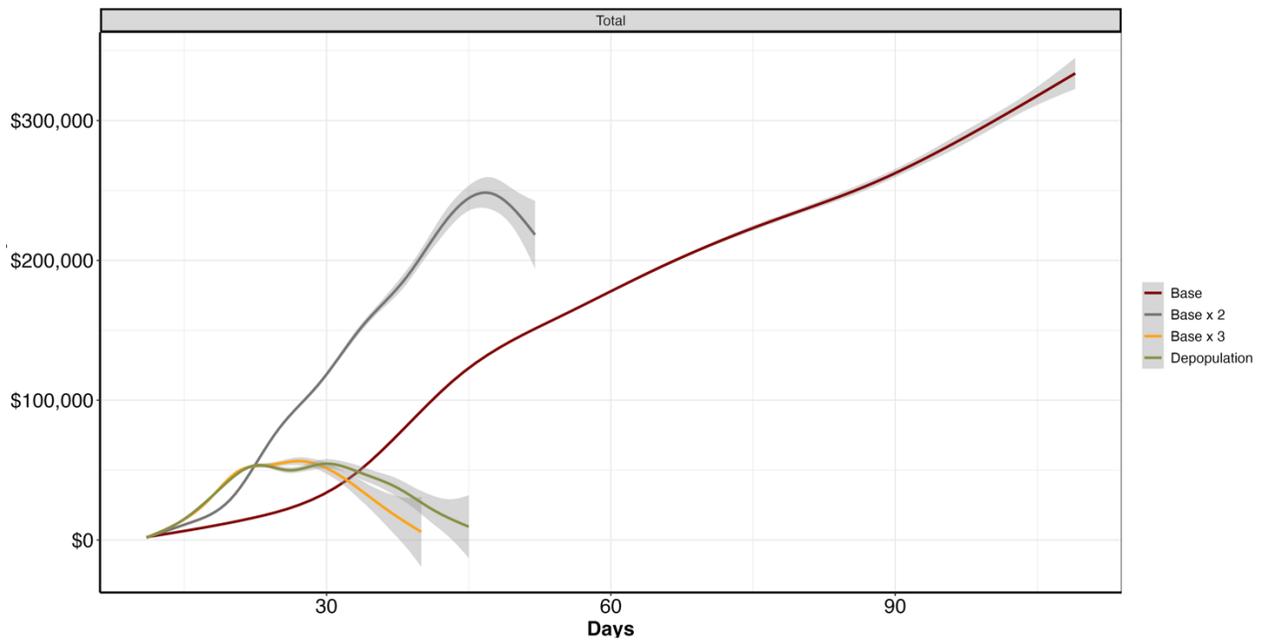

**Supplementary Material Figure S3.** Cost of the bullets used to cull by species for each scenario. The lines are fitted based on a Generalized additive model (GAM).

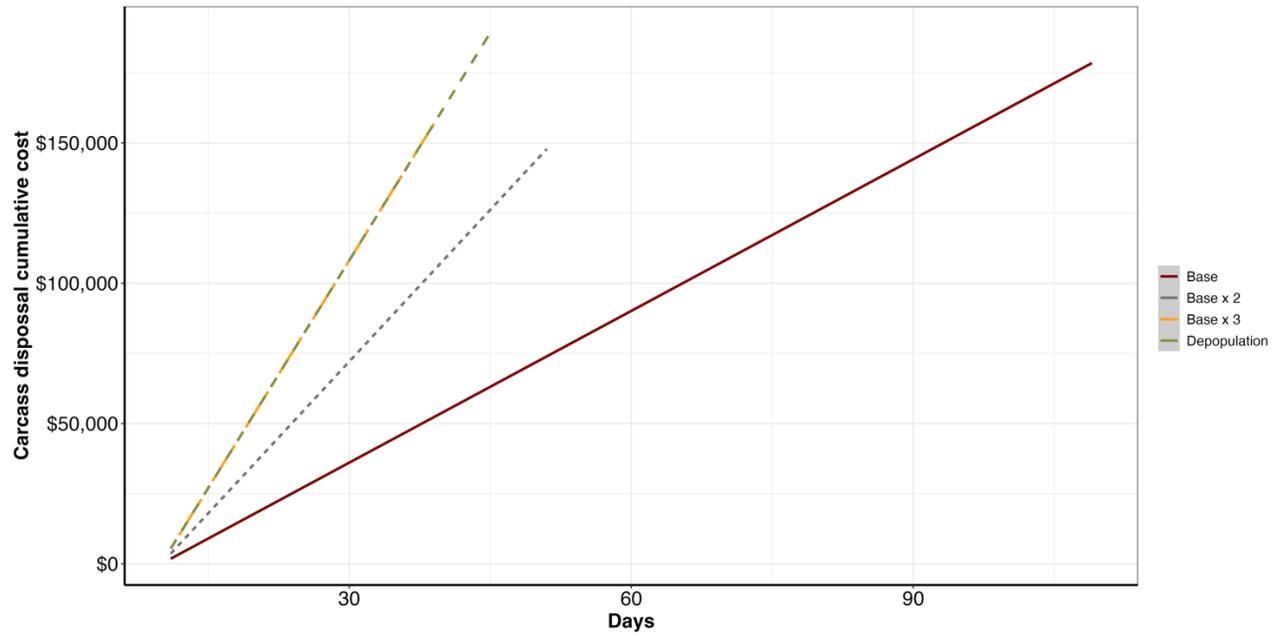

**Supplementary Material Figure S4.** Cumulative cost of the carcass disposal for each scenario.

The lines are fitted based on a Generalized additive model (GAM).

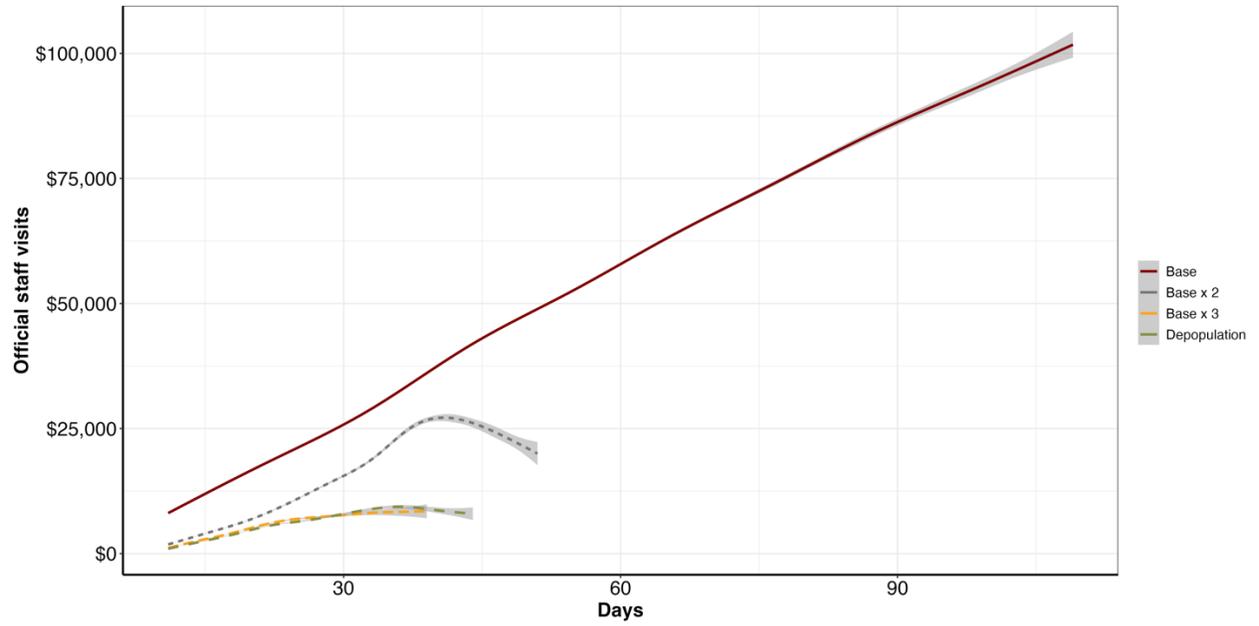

**Supplementary Material Figure S5.** Cumulative cost of the animal health official visits for each scenario. The lines are fitted based on a Generalized additive model (GAM).

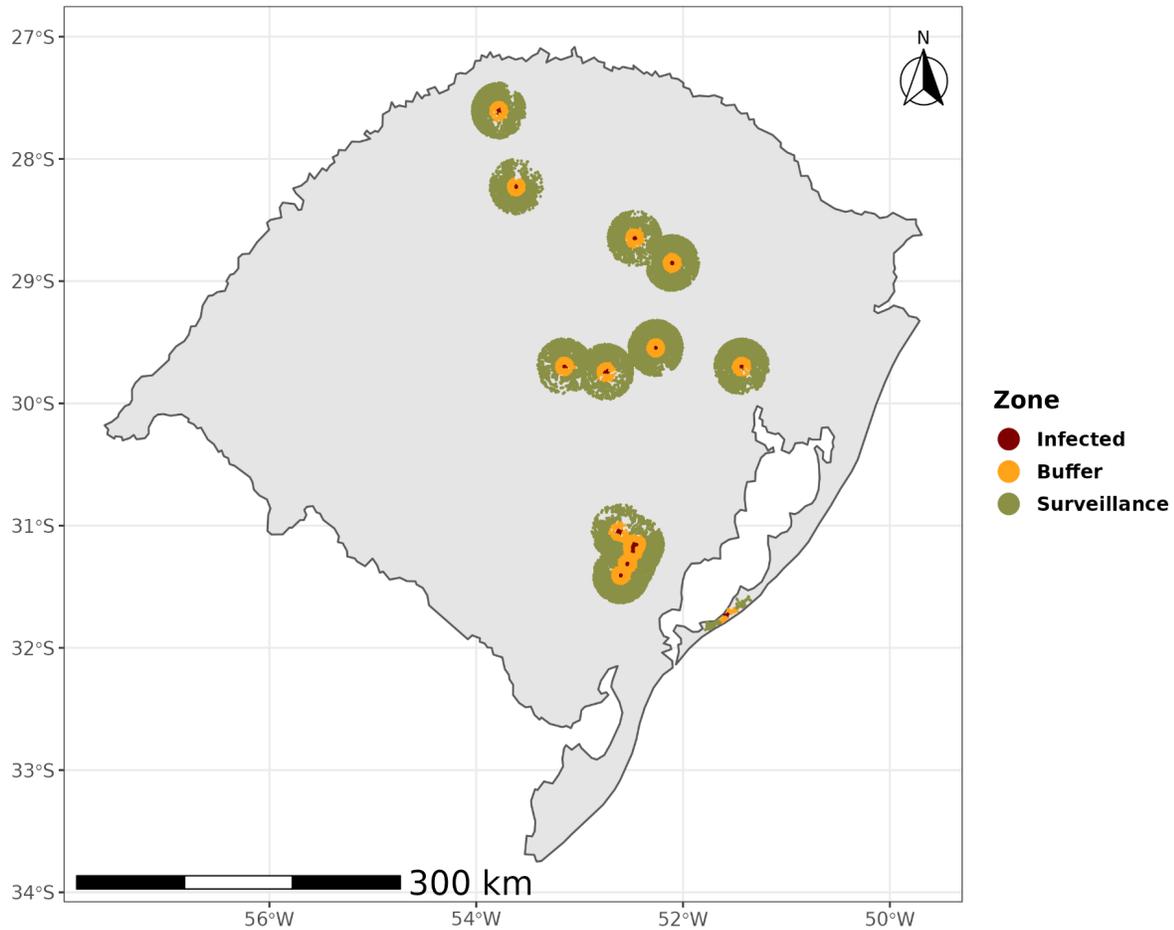

**Supplementary Material Figure S6**. **Control zones mapping.** The infected control zones of 3 km are represented as red dots, the 7 km buffer zone is defined as yellow dots, and the 15 km surveillance zone is described as green dots. Of note is that control zones with overlapping are merged.

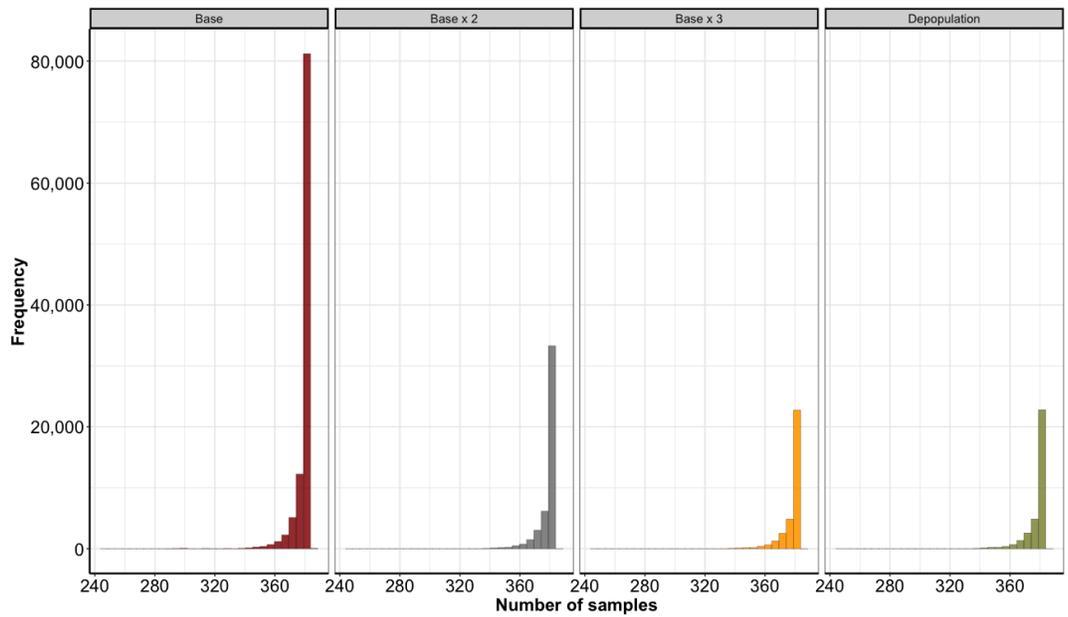

**Supplementary Material Figure S7**. Histogram distribution of the sampled sizes sampled within the control zone areas.

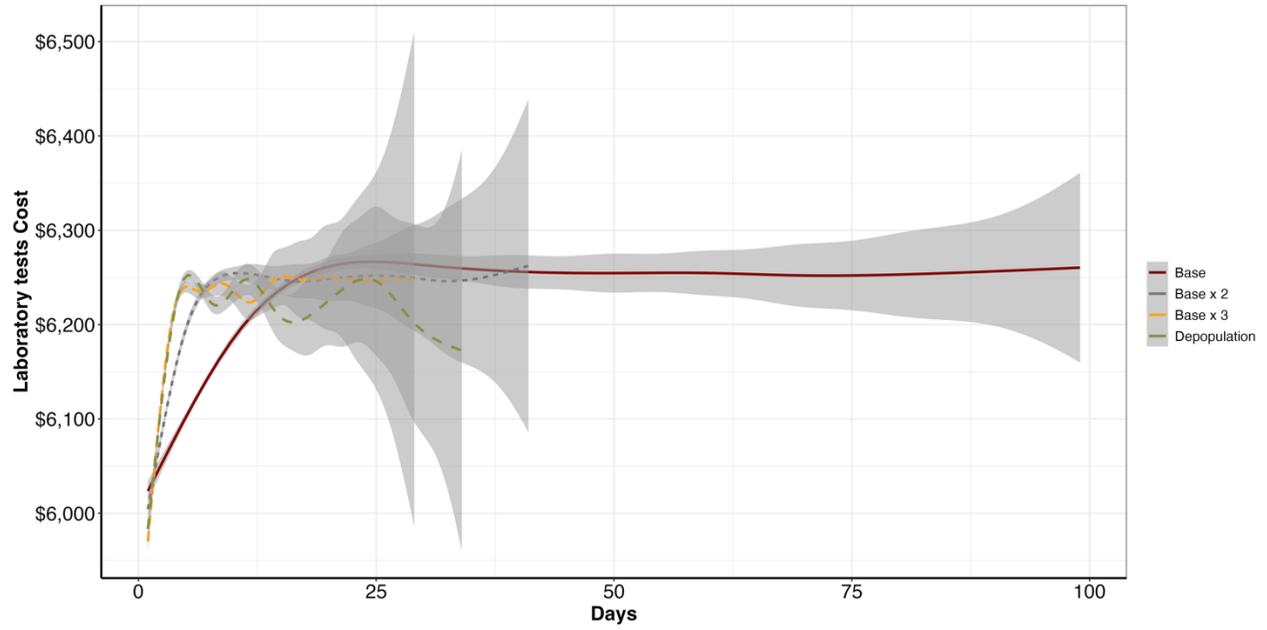

**Supplementary Material Figure S8.** Cost of the laboratory test for each scenario. The lines are fitted based on a Generalized additive model (GAM).

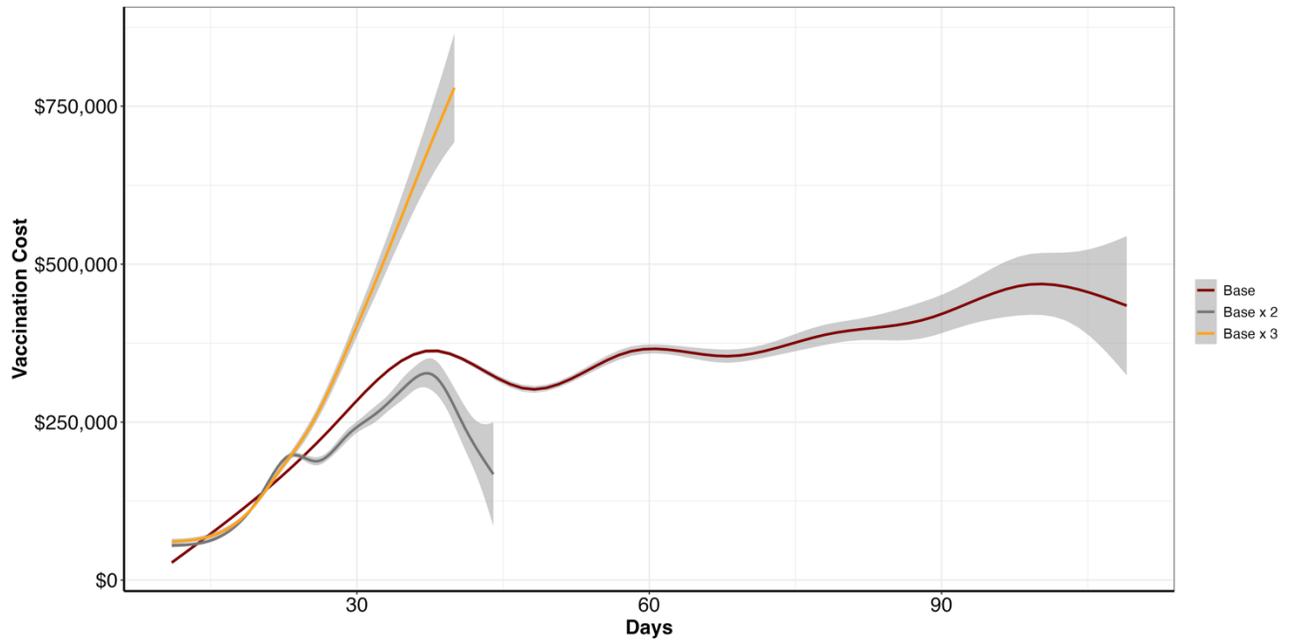

**Supplementary Material Figure S9.** Cost of the vaccination for each scenario. The lines are fitted based on a Generalized additive model (GAM).

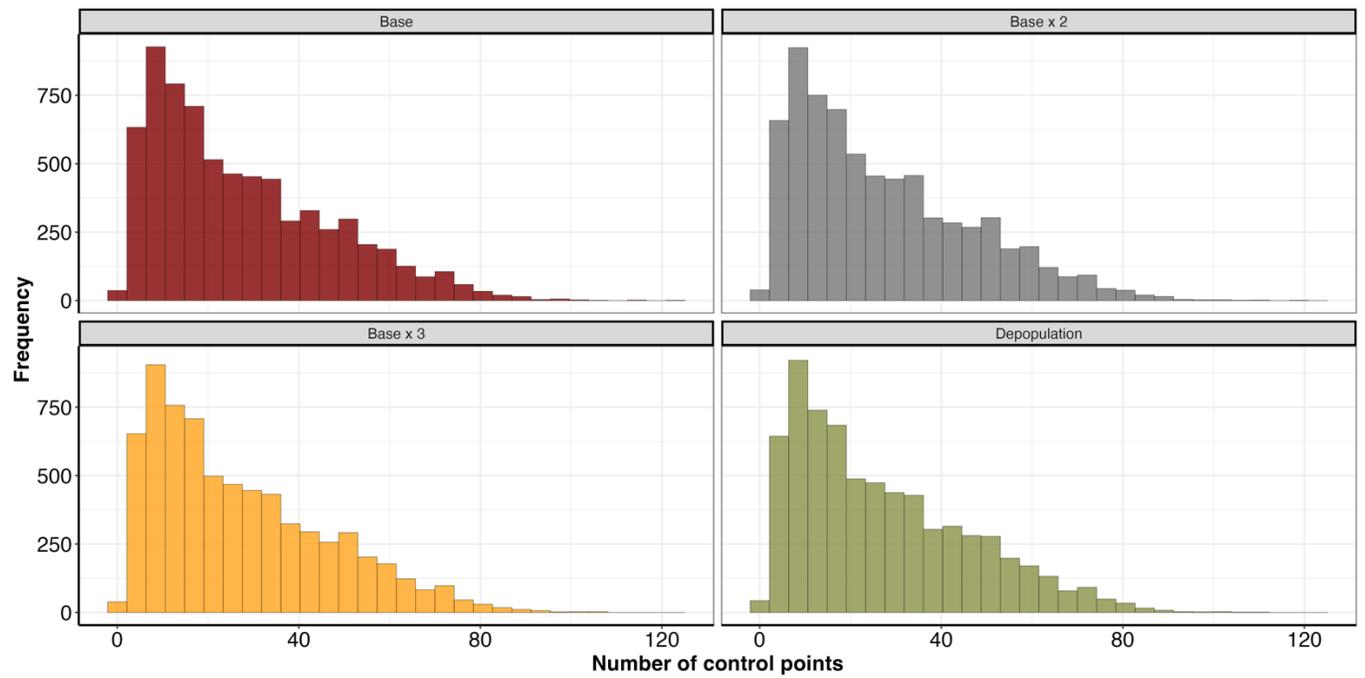

**Supplementary Material Figure S10.** Distribution of the number of control access points in each simulation by scenario.

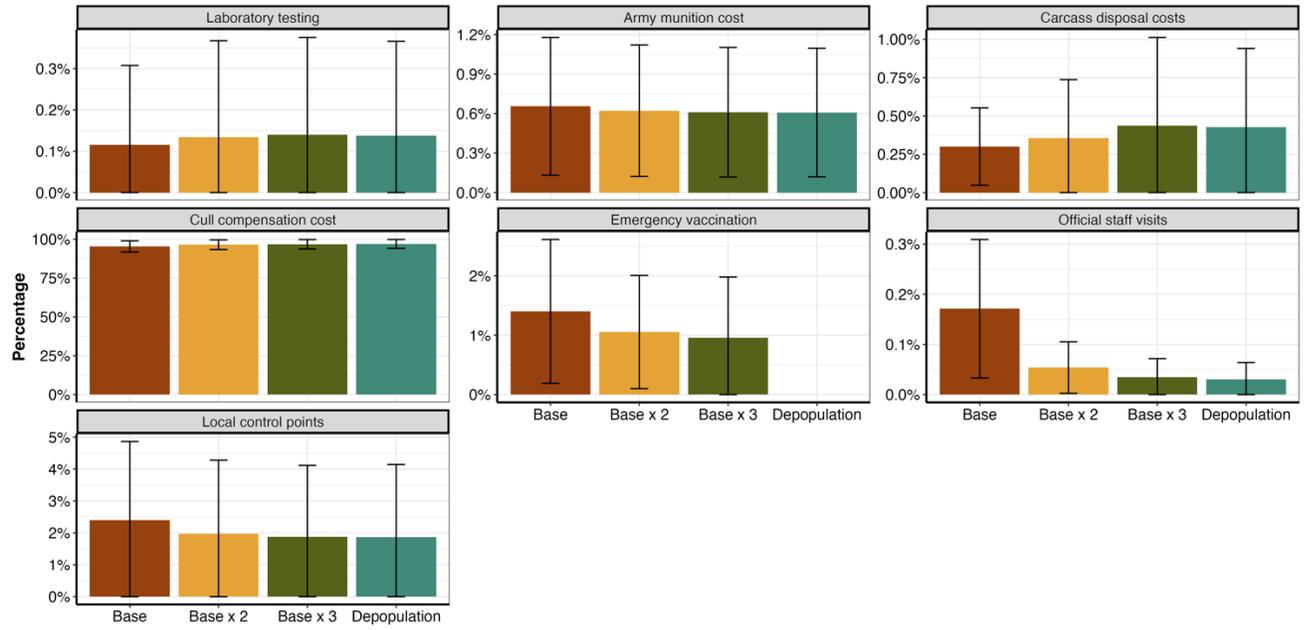

**Supplementary Material Figure S11. The itemized percentage cost by scenario.** The error bar represents the standard deviation, while the x-axis and color are the itemized costs. The Y-axis is in the log10 scale.

**Supplementary Material Table S1**. Parameters used to configure the simulation model in Rio Grande do Sul. The T stands for true, meaning the control action was applied. Otherwise, F represents that the parameters were not implemented.

| Parameter/Scenario | Base | Base x2 | Base x3 | Depopulation |
|---|---|---|---|---|
| **Initial Conditions** | | | | |
| Infected node | Selected randomly | Selected randomly | Selected randomly | Selected randomly |
| Initial day of the simulation | Selected randomly | Selected randomly | Selected randomly | Selected randomly |
| **Detection** | | | | |
| Days of control action | 120 | 120 | 120 | 120 |
| **Control Zones in kilometers** | | | | |
| Infected zone size | 3 | 3 | 3 | 3 |
| Buffer zone size | 5 | 5 | 5 | 5 |
| Surveillance zone size | 7 | 7 | 7 | 7 |
| **Movement Restriction** | | | | |
| Ban duration (days) | 30 | 30 | 30 | 30 |
| Infected zone movement | F | F | F | T |
| Buffer zone movement | F | F | F | T |
| Surveillance zone movement | T | T | T | T |
| Direct contacts movement | F | F | F | F |
| Traceback duration for movement | 1 | 1 | 1 | 2 |
| **Depopulation** | | | | |
| Limit of farms depopulated per day | 4 | 8 | 12 | 12 |
| Depopulation in the infected zone | F | F | F | F |
| Only depopulate infected farms | T | T | T | T |
| **Vaccination** | | | | |
| Days to achieve immunity | 15 | 15 | NA | NA |
| Farms vaccinated per day (infected zone) | 10 | 20 | 30 | 0 |
| Farms vaccinated per day (buffer zone) | 10 | 20 | 30 | 0 |
| Vaccination efficacy | 90% | 90% | 90% | NA |
| Vaccination of swine | F | F | F | NA |
| Vaccination of bovines | T | T | T | NA |
| Vaccination of small ruminants | F | F | F | NA |
| Vaccination in infected zone | T | T | T | NA |

| Vaccination in buffer zone | F | T | T | NA |
| --- | --- | --- | --- | --- |
| Vaccination delay (days) | 7 | 7 | 7 | NA |
| Vaccination in infectious farms | F | T | T | NA |

**Supplementary Material Table S2**. Cost of control measures across different scenarios. This table presents the median costs with various control measures by scenario, sorted by lower to higher cost.

| Scenario | Variable | Cost (median IQR and max) |
|---|---|---|
| Depopulation | Animal Health Official visits | $1,624 (IQR: $1,082 - $2,977, max: $31,931) |
| Base x3 | Animal Health Official visits | $1,894 (IQR: $1,082 - $3,247, max: $31,931) |
| Base x2 | Animal Health Official visits | $3,518 (IQR: $1,894 - $5,412, max: $43,567) |
| Base | Laboratory testing | $6,232 (IQR: $6,166 - $6,265, max: $6,298) |
| Base x2 | Laboratory testing | $6,232 (IQR: $6,166 - $6,265, max: $6,298) |
| Base x3 | Laboratory testing | $6,232 (IQR: $6,150 - $6,265, max: $6,298) |
| Depopulation | Laboratory testing | $6,232 (IQR: $6,150 - $6,265, max: $6,298) |
| Base | Animal Health Official visits | $12,989 (IQR: $7,036 - $20,836, max: $105,805) |
| Base | Animal Health Official visits | $21,633 (IQR: $12,619 - $34,252, max: $178,469) |
| Base x2 | Animal Health Official visits | $21,633 (IQR: $14,422 - $32,449, max: $147,823) |
| Base x3 | Carcass disposal | $27,041 (IQR: $16,224 - $37,857, max: $156,837) |
| Depopulation | Carcass disposal | $27,041 (IQR: $16,224 - $37,857, max: $189,286) |
| Base x3 | Traffic-control point | $94,744 (IQR: $42,471 - $178,796, max: $706,273) |
| Depopulation | Traffic-control point | $95,635 (IQR: $42,471 - $181,469, max: $668,257) |
| Base x3 | Emergency vaccination | $101,423 (IQR: $60,092 - $160,825, max: $890,199) |
| Base x2 | Traffic-control point | $104,545 (IQR: $44,550 - $208,496, max: $861,309) |
| Base x2 | Emergency vaccination | $110,167 (IQR: $63,850 - $175,319, max: $1,276,749) |
| Base | Emergency vaccination | $124,458 (IQR: $59,271 - $239,912, max: $1,417,456) |
| Base | Traffic-control point | $146,422 (IQR: $54,055 - $334,128, max: $1,853,299) |
| Base x3 | Depopulation | $1,507,160 (IQR: $760,291 - $2,836,677, max: $20,587,796) |
| Depopulation | Depopulation | $1,541,582 (IQR: $780,099 - $2,907,191, max: $52,275,811) |
| Base x2 | Depopulation | $1,609,565 (IQR: $814,810 - $2,945,982, max: $20,655,679) |
| Base | Depopulation | $1,765,022 (IQR: $901,384 - $3,058,477, max: $21,478,453) |

**Supplementary Table S2**. Proportion of control measures across different scenarios. The data has median, interquartile ranges (IQR), and maximum values.

| Scenario | Variable | Proportion (median IQR and max) |
|---|---|---|
| Base | Animal Health Official visits | 0.01 (IQR: 0 - 0.01, max: 0.04) |
| Base | Laboratory testing | 0 (IQR: 0 - 0.01, max: 0.09) |
| Base | Carcass disposal | 0.01 (IQR: 0.01 - 0.02, max: 0.09) |
| Base | Emergency vaccination | 0.05 (IQR: 0.03 - 0.09, max: 0.34) |
| Base | Traffic-control point | 0.08 (IQR: 0.04 - 0.14, max: 0.61) |
| Base | Depopulation | 0.86 (IQR: 0.76 - 0.92, max: 1) |
| Base x 2 | Animal Health Official visits | 0 (IQR: 0 - 0, max: 0.02) |
| Base x 2 | Laboratory testing | 0 (IQR: 0 - 0.01, max: 0.13) |
| Base x 2 | Carcass disposal | 0.01 (IQR: 0.01 - 0.02, max: 0.17) |
| Base x 2 | Emergency vaccination | 0.04 (IQR: 0.02 - 0.07, max: 0.4) |
| Base x 2 | Traffic-control point | 0.06 (IQR: 0.03 - 0.12, max: 0.65) |
| Base x 2 | Depopulation | 0.9 (IQR: 0.82 - 0.95, max: 1) |
| Base x 3 | Animal Health Official visits | 0 (IQR: 0 - 0, max: 0.01) |
| Base x 3 | Laboratory testing | 0 (IQR: 0 - 0.01, max: 0.1) |
| Base x 3 | Carcass disposal | 0.01 (IQR: 0.01 - 0.02, max: 0.31) |
| Base x 3 | Emergency vaccination | 0.03 (IQR: 0.02 - 0.06, max: 0.35) |
| Base x 3 | Traffic-control point | 0.06 (IQR: 0.03 - 0.11, max: 0.58) |
| Base x 3 | Depopulation | 0.91 (IQR: 0.84 - 0.95, max: 1) |
| Depopulation | Animal Health Official visits | 0 (IQR: 0 - 0, max: 0.02) |
| Depopulation | Laboratory testing | 0 (IQR: 0 - 0.01, max: 0.1) |
| Depopulation | Carcass disposal | 0.01 (IQR: 0.01 - 0.02, max: 0.28) |
| Depopulation | Traffic-control point | 0.06 (IQR: 0.03 - 0.11, max: 0.66) |
| Depopulation | Depopulation | 0.92 (IQR: 0.85 - 0.96, max: 1) |